\documentclass[reprint, secnumarabic, amssymb, amsmath, nobibnotes, 
     aps, prl, graphicx]{revtex4-1}
\usepackage{graphicx}		
\begin{document}
\title{New method of solving the many-body Schr\"{o}dinger equation.}
\author{V.M.Tapilin}
\email{tapilin@catalysis.ru}
\affiliation {Boreskov Institute of Catalysis, Novosibirsk 630090, Russia}
\date{\today}
\begin{abstract}
A method of solving the Schr\"{o}dinger equation based on the use of constant 
particle-particle interaction potential surfaces (IPS) is proposed. The many-body 
wave function is presented in a configuration interaction form, with coefficients 
depending on the total interaction potential. The corresponding set of linear 
ordinary differential equations for the coefficients was developed. To reduce 
the computational work, a hierarchy of approximations based on interaction 
potential surfaces of a part of the particle system was worked out. 
The solution of a simple exactly solvable model and He-like ions proves that 
this method is more accurate than the conventional configuration interaction 
method and demonstrates a better convergence with a basis set increase.
\end{abstract}
\maketitle

\paragraph*{Introduction.---}M{\o}ller-Plesset perturbation theory and configuration 
interactions (CI) are the conventional methods of treating electron-electron correlation 
in the theory of atoms and molecules \cite{sherrill:110902}. Unfortunately, due to the 
presence of the correlation cusp \cite{kato,King} in the wave function, both of them 
reveal slow convergence of electron energy with basis set increasing. 

Density functional theory (DFT) \cite{Hohen, Kohn1965, Kohn1996, Chris} is another 
approach for the solution of a quantum many-body problem. Based on solving Kohn-Sham 
equations \cite{Kohn1965}, it has been successfully applied to many problems \cite{Chris}. 
Unfortunately, the exact form of this functional is still unknown, and its approximated 
forms do not always provide needed accuracy, for example, in treating the systems with 
strong electron-electron correlations \cite{Ivan, PhysRevB.74.125120, Dagotto08072005}. 

All these arguments give reasons to search for other ways of treating the correlation problem. 
To speed up the convergence, explicitly correlated have been developed over 
the last two decades \cite{ Yousaf2009303,KlopperInt,doi:10.1021/cr200168z} in which 
the wave function explicitly depends on electron-electron spacing. Iterative complement 
interaction method has been formulated in \cite{PhysRevLett.93.030403,PhysRevA.72.062110}. 

This paper is aimed at developing another way of treating the correlation problem presented 
in \cite{TapilinJSC,*TapilinAR}. The theory is based on the introduction of constant 
particle-particle interaction potential surfaces (IPS). From the definition of such surfaces it 
follows that particle-particle interaction acts along the normal to the surface and, therefore, 
does not influence particle motion on the surface. Further a new form of many-body wave 
function and equations to find it will be proposed and applied to a simple model system 
and He-like ions. 

\paragraph*{Configuration weight functions and equations determining them.---}Consider the 
Schr\"{o}dinger equation of $n$ interacting particles and
introduce a collective variable
\begin{equation}
	\frac{1}{p(\mathbf{R})}=V(\mathbf{R})=
\sum_{i=1}^{n-1}\sum_{j>i}^{n}\frac{1}{r_{ij}}=
\sum_{i=1}^{n-1}\sum_{j>i}^{n}v_{ij}.\label{vint} 
\end{equation}
Here $\mathbf{R}$ stands for a set of particle coordinates $\mathbf{r}_1,...,\mathbf{r}_n$, 
$r_{ij}=|\mathbf{r}_{i}-\mathbf{r}_{j}|$. 

A constant IPS $V(\mathbf{R})=1/p$ selects a subspace of particle 
coordinates in which particle motion is correlated \textit{ab origin} due to the demand 
remaining at the surface rather than particle interaction. The resulting interaction force 
does not act on particle movement along the surface, therefore, the movement can be described 
by a function $\Phi_{\mathbf{i}}(\mathbf{R})=\phi_{i_1}(\mathbf{r}_1)\ldots\phi_{i_n}(\mathbf{r}_n)$
satisfying to the Schr\"{o}dinger equation of non-interacting particles $H(\mathbf{R})$.
Here we introduced vectors $\mathbf{i}$ with components $i_1,\ldots,i_n$.
To satisfy the Schr\"{o}dinger equation of interacting particles 
we represented the wave function in the form
\begin{equation}\label{Psi}
	\Psi(\mathbf{R})=\sum_{\mathbf{i}}\chi_{\mathbf{i}}(p(\mathbf{R}))\Phi_{\mathbf{i}}(\mathbf{R}).
\end{equation}
Function (\ref{Psi}) has the form of CI function in which constant coefficients 
are replaced by functions $\chi_{\mathbf{i}}(p(\mathbf{R}))$ depending on interaction potential 
at points $\mathbf{R}$.

Energy minimization in respect to $\chi_{\mathbf{i}}$ leads to equations
\begin{equation}\label{eqc}
\begin{split}
	\sum_\mathbf{j}&\left[-\frac{t_\mathbf{ij}(p)}{2}\frac{d^2\chi_{\mathbf{j}}(p)}{dp^2}
	-\left(\frac{t_\mathbf{ij}(p)}{p}+\frac{u_\mathbf{ij}(p)}{2}\right)\frac{d\chi_{\mathbf{j}}(p)}{dp}\right.\\ &\left.+
  \left(h_\mathbf{ij}(p)+\frac{1}
{p}\right)\chi_{\mathbf{j}}(p)\right]=E\sum_\mathbf{j}s_\mathbf{ij}(p) \chi_{\mathbf{j}}(p),
\end{split}
\end{equation}
where
\begin{align}
	 \label{T} t_\mathbf{ij}(p)&=\langle	\Phi_\mathbf{i}
	(\mathbf{R})|(\nabla_\mathbf{R} p)^2|\Phi_\mathbf{j}(\mathbf{R})\rangle_p, \\
	 \label{U} u_\mathbf{ij}(p)&=\langle	\Phi_\mathbf{i}
	(\mathbf{R})|\nabla_\mathbf{R} p\nabla_\mathbf{R}|\Phi_\mathbf{j}(\mathbf{R})\rangle_p ,\\
	 \label{H} h_\mathbf{ij}(p)&=\langle	\Phi_\mathbf{i}(\mathbf{R})|H(\mathbf{R})
	|\Phi_\mathbf{j}(\mathbf{R})\rangle_p \\ 
	 \label{S} s_\mathbf{ij}(p)&=\langle	\Phi_\mathbf{i}(\mathbf{R})|
	\Phi_\mathbf{j}(\mathbf{R})\rangle_p,  
	\end{align}
 
Eq. (\ref{eqc}) is a set of linear ordinary differential equation with eigenvalues
equal to the system energy. The terms containing derivatives of $\chi$ describe 
additional contributions to kinetic energy arising when redistribution 
of electrons between different interaction potential surfaces occurs. There is no 
such redistribution for non-interacting particles. For this case functions $\chi$ are
constant and differential equations (\ref{eqc}) reduce to the Schr\"{o}dinger equations 
of non-interacting particles. Beside electron redistribution between different $p$, 
functions $\chi$ determine the contributions of different configurations for constant $p$, 
hereinafter referred to as \textit{configuration weight function}.

The boundary conditions for $\chi$ follow from the demand for $\Psi$ to be finite in the 
whole space $\chi_i(p)s_{ij}(p)\chi_j(p)<\infty$. For small $p$ Eqs.(\ref{eqc}) can be 
approximated by 	$2d\chi/dp+\chi/Z=0$ with solution $\chi(p)\approx e^{Zp/2}$ which 
provides the boundary conditions for solutions of (\ref{eqc}). Determine functions 
$\omega_{ij}^l(p,E)$ which are solutions of (\ref{eqc}) for energy $E$  with different 
boundary conditions $	\omega_{ij}^l(0)=\delta_{ij}$, and $d\omega_{ij}^l(0)/dp=Z\delta_{ij}/2$.
The common solution of (\ref{eqc}) can be presented in the form
$\chi_i^l(p,E)=\sum_k c_k^l\omega_{ki}^l(p,E)$.

 Since particle-particle interaction tends
to separate the particles, it can be expected that $\chi(p)$ is a growing function 
of $p$, with the growth being not too fast to prevent total function divergence.   
The valid boundary conditions depend on the behavior of $t_{\mathbf{ij}}$ and  
$u_{\mathbf{ij}}$ when $p\to\infty$ and should be discussed for particular problems.  

\paragraph*{The particle-particle IPS.---}The constant particle-particle IPS
is a plane in the space of the pair potentials $v_{ij}=1/r_{ij}$, 
which will be refer to as \textit{v-space}. The dimensionality of v-space is $n(n-1)/2$. Each 
point of v-space determines the relative particle positions $r_{ij}$ in the usual space, which 
we call \textit{r-space}, so the set of $\mathbf{R}$ belonging to the same surface can be easily 
determined. Not all of $r_{ij}$ are independent because all of them are determined by $3(n-1)$ 
particles coordinates placing the 1st particle at the coordinate system origin. Due to the 
multidimensionality of the plane and a dependence of integration limits of a particle on position 
of the other particles the numerical integration over the surface can be performed only for several 
particle systems. It means that the developed theory can be applied at the best only for such 
systems, and an extension of the theory to bigger systems needs to be simplified. Possible 
simplifications are proposed below.

At first, one can introduce a set of approximations to the theory 
based on the lowering the dimension of IPS by averaging over the coordinates of $m$ particles. 
Correspondent surfaces will be denoted $S_m$, the dimension of such IPS is $3(n-m-1)$. 
Averaging over all particles but two describes the motion of exactly correlating particle pairs in the 
middle field of other particles can be called \textit{independent pair approximation}. The same way can 
be introduced independent \textit{triplet}, \textit{quadruple}, etc. approximations. Introducing
IPS for poitential acting on a particle from the other ones, each $S_m$ IPS can be reduced to $S_{1m}$
with dimension $n-m-1$. For a particle at $\mathbf{r}_1$ it consists of $n-m-1$ 
spheres of radius $r_{1i}$. 

\paragraph*{A simple exactly solvable model.}To test the developing theory, we considered a simple model, 
three particles in one dimensional infinite square potential well, and solved the Scr\"{o}dinger equation 
directly, with configuration interaction method and with different approximations of the developing theory. 
To avoid errors in derivatives approximation by finite differences we \textit{ab origin} used the discrete 
space  containing ten points together with the border ones in which kinetic energy operator acting
on a particle at point $i$ is $2\delta_{i_1j_1}-\delta_{i_1,j_1\pm 1}$, and the interaction  
between a pair of particles $v_{i_1j_2}=1/(|i_1-j_2|+\lambda)$ with $\lambda = 0.1$ to prevent 
$v$ from becoming infinity at $i_1=j_2$. The total matrix of 512x512 order constructed from these matrix 
elements for three particles has been diagonalized. 

A complete orthonormal basis functions set turn into zero at the border points is
\begin{equation}\label{1f}
\varphi_\alpha(i)=\sqrt{\frac{2}{\pi}}\sin\frac{\alpha i\pi}{9},\:i=0,\ldots,9,\:\alpha =1,\ldots,8,
\end{equation}
where $\alpha$ numerates functions and $i$ numerates points. From these functions a set of 
configurations can be constructed $\Phi_\alpha(\mathbf{i})=\varphi_{\alpha_1}(i_1)\varphi_{\alpha_2}(i_2)
\varphi_{\alpha_3}(i_3)$. To check the convergence of CI method the problem has been 
solved for different number $n_f$ of functions (\ref{1f}) used to   
configuration construction. The four lowest eigenvalues for different $n_f$ are shown 
in Table \ref{tab:exite}, columns $CI$.

The IPS has been constructed for the total interaction, $S_3$, for the
potential $v_{12}+v_{13}$, $S_{13}$, and for $v_{12}$, $S_2$. On some surfaces functions (\ref{1f})
are linear dependent and for such surfaces a new orthonormal set of one-particle functions 
has been constructed. The matrix corresponding to $\chi$ has been obtained and diagonalized.  
The four lowest eigenvalues for different $n_f$ are shown 
in Table \ref{tab:exite} in columns $S_3$, $S_{13}$ and $S_2$

\begin{table}
\caption{Convergence with basis set increase for ground (g) and
	1st, 2nd, and 3rd excited states obtained with $S_3$, 
	$S_{13}$, $S_2$ and $CI$ matrices.}
	\centering
		\begin{tabular}{cccccc}\hline\hline 
$n_f$&state & $S_3$    & $S_{13}$    & $S_2$ &  $CI$\\ 
2&g    &  2.5167521 &  2.5523628 & 3.2802597&  4.7233508 \\
 &1st  &  2.6007468 &  2.5630105 & 3.3433421&  4.7233508 \\
 &2nd  &  2.6007468 &  2.5810447 & 3.5953573&  4.9761493 \\
 &3rd  &  2.8993370 &  2.7509650 & 3.7269786&  4.9761493 \\
3&g    &  2.5164929 &  2.5164929 & 2.6297927&  2.7116594 \\
 &1st  &  2.5624931 &  2.5512484 & 2.6779104&  3.1982957 \\
 &2nd  &  2.5624931 &  2.5513833 & 2.6926681&  3.1982957 \\
 &3rd  &  2.6311803 &  2.6246885 & 2.7596655&  3.5842098 \\
4&g    &  2.5164929 &  2.5164929 & 2.5796667&  2.6802123 \\
 &1st  &  2.5512299 &  2.5512296 & 2.5965684&  2.6803017 \\
 &2nd  &  2.5512299 &  2.5512296 & 2.6583777&  2.6803017 \\
 &3rd  &  2.6246904 &  2.6246885 & 2.6678403&  2.7694084 \\
5&g    &  2.5164929 &  2.5164929 & 2.5560362&  2.6101078 \\
 &1st  &  2.5512296 &  2.5512296 & 2.5705697&  2.6273272 \\
 &2nd  &  2.5512296 &  2.5512296 & 2.6028451&  2.6273272 \\
 &3rd  &  2.6246885 &  2.6246885 & 2.6369918&  2.6596377 \\
6&g    &  2.5164929 &  2.5164929 & 2.5434281&  2.5722193  \\
 &1st  &  2.5512296 &  2.5512296 & 2.5623772&  2.5938677 \\
 &2nd  &  2.5512296 &  2.5512296 & 2.5847363&  2.5938677 \\
 &3rd  &  2.6246885 &  2.6246885 & 2.6308173&  2.6415573 \\
7&g    &  2.5164929 &  2.5164929 & 2.5304009&  2.5478066 \\
 &1st  &  2.5512296 &  2.5512296 & 2.5569004&  2.5751546 \\
 &2nd  &  2.5512296 &  2.5512296 & 2.5676064&  2.5751546\\
 &3rd  &  2.6246885 &  2.6246885 & 2.6269072&  2.6336853 \\
8&g    &  2.5164929 &  2.5164929 & 2.5164929&  2.51649929\\ 
 &1st  &  2.5512296 &  2.5512296 & 2.5512296&  2.55122966 \\
 &2nd  &  2.5512296 &  2.5512296 & 2.5512296&  2.55122966 \\
 &3rd  &  2.6246885 &  2.6246885 & 2.6246885&  2.62468855 \\
\hline \hline
			\end{tabular}
	\label{tab:exite}
\end{table}
The results show that all of the applied methods for $n_f=8$ give exactly the same 
results for the ground and exited states. This situation continues in $S_3$ and $S_{13}$ 
cases up to $n_f=4$ for all states, and up to $n_f=3$ for the ground state in spite of 
one-body basis set reduction. The result is a sequent that up to $n_f=4$ the number
of linear independent functions constructed with (\ref{1f}) remains unchanged.
The basis set decrease leads to the decrease of linear independent functions,  and the 
accuracy of calculations drops significantly, faster for CI method.

\paragraph*{He-like ions.}Solving the Schr\"{o}dinger equations for He-like ions 
it is convenient to use for length and energy corresponding atomic units divided 
by nuclear charge $Z$ and $Z^2$, respectively. 
For description of $^1S$ states we used 
$2e^{-r}$, $(1-r/2)e^{-r/2}/\sqrt{2}$ and $2(1-2r/3+2r^2/27)e^{-r/3}/\sqrt{27}$
wave functions corresponding to $1s$, $2s$ and $3s$ states of an a electron in the
nuclear field. From these function 3 configuration with the lowest 
energy, $\Phi_1(r_1,r_2)=\phi_1(r_1)\phi_1(r_2)$, $\Phi_2(r_1,r_2)=(\phi_1(r_1)
\phi_2(r_2)+\phi_1(r_2)\phi_2(r_1))/\sqrt{2}$ and $\Phi_3(r_1,r_2)=(\phi_1(r_1)
\phi_3(r_2)+\phi_1(r_2)\phi_3(r_1))/\sqrt{2}$, have been constructed.  
Matrix elements between these functions can be obtained analytically. 

To define the boundary condition when $p\to\infty$ we represent
the solution of ($\ref{eqc}$) at a point $p$ as $e^{\lambda p}$. 
Substitution of this representation in ($\ref{eqc}$) leads to
\begin{equation}\label{eqp}
\begin{split}
	\sum_{j=1}^n\left[-\lambda^2s_{ij}(p)
	-\lambda ((2s_{ij}(p)/p+u_{ij}(p))+h_{ij}(p)\right.\\ \left.
	+s_{ij}(p)/Zp-Es_{ij}(p)\right]\chi_j(p)=0,\ i=1,...,n,
\end{split}
\end{equation}
$n$ is the number of configuration taking into account.
 Set (\ref{eqp}) has non-zero solution if $\det(\Lambda)=0$
where matrix $\Lambda$ is determined by the expressions in the square brackets
of (\ref{eqp}). Determine functions $\omega_{ij}^r(p,E)$
which are solutions of (\ref{eqc}) for energy $E$ and satisfy boundary conditions
	$\omega_{ij}^r(p,E)=\delta_{ij}$, and $d\omega_{ij}^r(p,E)/dp=\lambda(E)\delta_{ij}$,
where $\lambda$ is a root of $\det(\Lambda)$ satisfying condition $e^{2\lambda p}
s_{ij}(p)<\infty$.
The common solution of (\ref{eqc}) can be presented in the form
$	\chi_i^r(p,E)=\sum_{k}^n c_k^r\omega_{ki}^r(p,E)$. Coefficients $c^l$, $c^r$ and
energy $E$ are determined from the demand that functions 
$\chi_i^l$ must continuously pass to functions $\chi_i^r$ at a matching point $p$
together with their 1st derivatives. 

\begin{table*}[t]
\vspace{5mm}
TABLE II The ground states energies of He-like ions. 
\begin{center}
\begin{tabular}{cccccccc} \hline \hline
Ion&\multicolumn{7}{c}{ Energy, a.u.} \\
     &HF$^a$ &1&2&3& CI$^b$ & CI$^c$& Exp.$^d$  \\ 
   \hline
$H^{-}$    &          &  -0.498665 &  -0.527171 &  -0.527768& -0.52760& -0.5277303& \\
He         &- 2.86171 &  -2.880042 &  -2.901782 &  -2.903896& -2.90325& -2.9037236& -2.9038\\                    
$Li^{+}$   & -7.23633 &  -7.257493 &  -7.278072 &  -7.280982& -7.27928&  -7.279819& -7.2804\\                  
$Be^{2+}$  &-13.61130 & -13.633947 & -13.654148 & -13.657667&-13.65485& -13.655551& -13.6574\\                 
$B^{3+}$   &-21.98607 & -22.010000 & -22.030095 & -22.034159&-22.03020& -22.030875& -22.0360\\                   
$C^{4+}$   &-32.36137 & -32.385852 & -32.405967 & -32.41054 &-32.40544& -32.406070&  -32.4174\\   
$N^{5+}$   &-44.73618 & -44.761590 & -44.781795 & -44.786878&-44.78061& -44.781141& -44.8035\\ 
$O^{6+}$   &-59.11159 & -59.137256 & -59.157592 & -59.163168&-59.15574& -59.156222& -59.1958\\ 
$F^{7+}$   &-75.48702 & -75.512875 & -75.533370 & -75.539432& -75.5308& -75.531401& -75.5970\\    
$Ne^{8+}$  &-93.86174 & -93.888460 & -93.909133 & -93.915678&-93.90592& -93.906452& -94.0086\\  
$Na^{9+}$  &          &-114.264020 &-114.284885 &-114.291910&         &-114.28165 &\\
$Mg^{10+}$ &          &-136.639562 &-136.660629 &-136.668132&         &-136.65672 &\\                    
$Al^{11+}$ &          &-161.015090 &-161.036367 &-161.044346&         &-161.03180 &\\                  
$Si^{12+}$ &          &-187.390608 &-187.412100 &-187.420554&         &-187.40687 &\\                 
$P^{13+}$  &          &-215.766116 &-215.787829 &-215.796757&         &-215.78191 &\\                   
$S^{14+}$  &          &-246.141617 &-246.163554 &-246.172956&         &-246.15697 &\\   
$Cl^{15+}$ &          &-278.517113 &-278.539277 &-278.549152&         &-278.53201 &\\ 
$Ar^{16+}$ &          &-312.892603 &-312.914998 &-312.925344&         &-312.90704 &\\ 
\hline \hline                                     
$^a$ Ref.\cite{Clementi}. \\  
$^b$ Ref.\cite{Weiss}. \\  
$^c$ Ref.\cite{Saha}  \\                                 
$^d$ Ref.\cite{Moore,Bashkin}                                   
\end{tabular}                                     
\end{center}  
\end{table*}   
To solve (\ref{eqc}) 
the Runge-Kutta 4th-order method was employed. The energies obtained for the 
ground states of He-like ions are shown in Tables II together with HF and CI results.
The use of only one-configuration approximation gives energies slightly below 
Hartree-Fock limit. Inclusion of the second configuration give results comparable
with those of CI with 35 configurations \cite{Weiss}. Our results are slightly above 
the CI results from $H^-$ up to $B^{3+}$ and below CI results for the rest of 
calculated ions. Inclusion of the third configuration gives the lowest energies
presented in the table. More over the energies of ions up to $Be^{2+}$  turn out to be
below experimental values. However, accounting mass correction factors for
these ions $M/(M+m)$, gives energies for $He$ -2.903501, for $Li^+$ -7.280415 
and $Be^{2+}$ -13.656841.

The configuration weight functions for 1-, 2- and 3-configuration 
approximation are shown in Fig. 1, Fig. 2, and Fig. 3 and Fig. 4, correspondingly.  
\begin{figure}[h]
\label{wf1}
\begin{center}
\scalebox{0.35}{\includegraphics{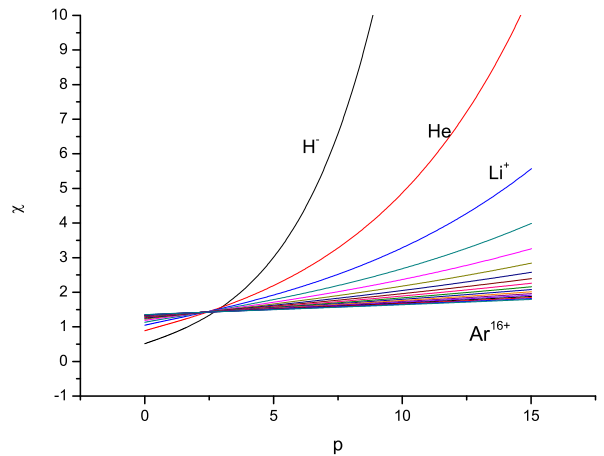}}
\caption{$1s1s$ configuration weight functions for $H^-$,...,$Ar^{16+}$ in 1-configuration  
         approximation.}
\end{center}
\end{figure}

\begin{figure}
\label{wf2}
\begin{center}
\scalebox{0.35}{\includegraphics{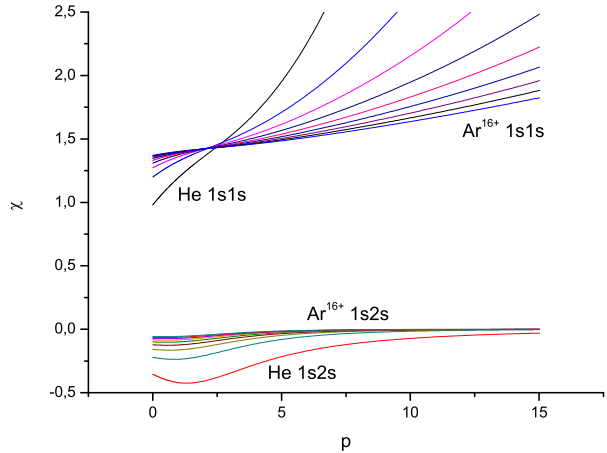}}
\caption{$1s1s$ and $1s2s$ configuration weight functions from $He$ to 
          $Ar^{16+}$ for even atomic numbers in 2-configuration approximation.} 
\end{center}
\end{figure}

\begin{figure}[t]
\label{wf3_He}
\begin{center}
\scalebox{0.35}{\includegraphics{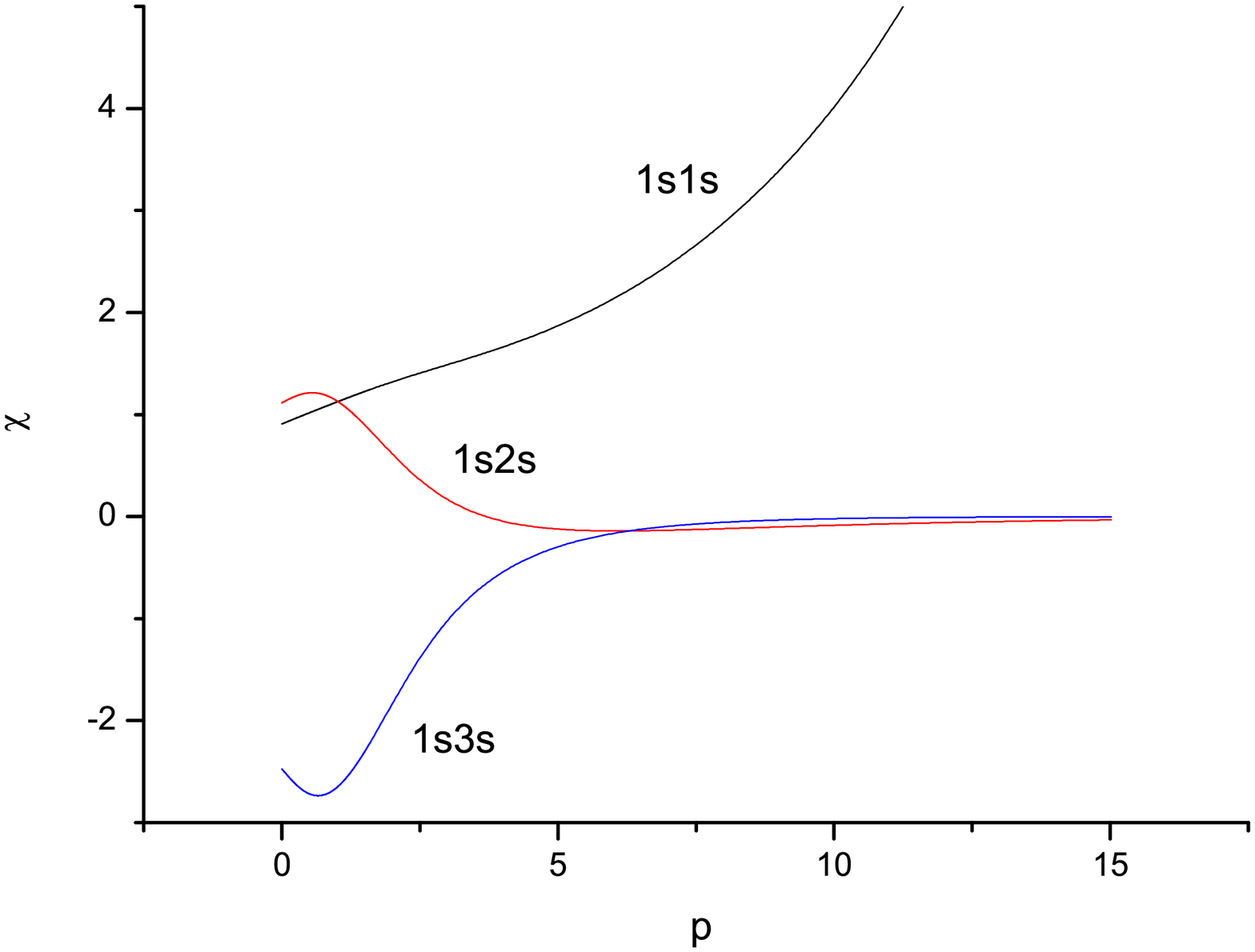}}
\caption{$1s1s$, $1s2s$ and $1s3s$ configuration weight functions for $He$
                         in 3-configuration approximation. }
\scalebox{0.35}{\includegraphics{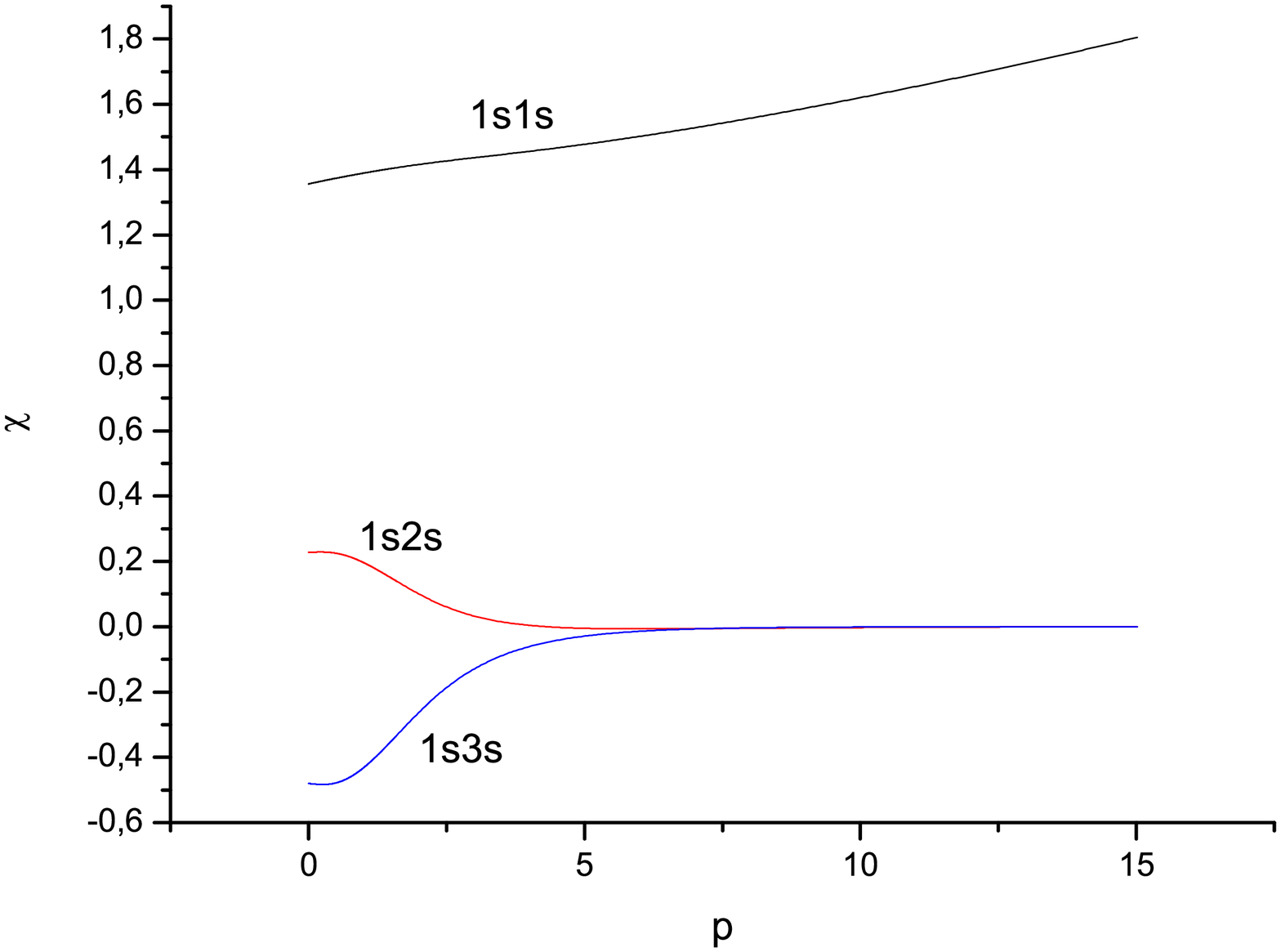}}
\caption{$1s1s$, $1s2s$ and $1s3s$ configuration weight functions for $Ar^{16+}$
in 3-configuration approximation}
\end{center}
\end{figure}

In all cases 
$1s1s$ configuration weight functions are increasing functions. The functions growth 
slows down with increasing nuclear charges and tends to be constant. The growing 
weight function decreases the probability of finding an electron at a small separation which 
increases for a bigger separation in comparison with a non-interacting case.
For 2-configuration approximation $1s2s$ functions have noticeable values at small $p$ 
decreasing with the growth of $p$ and atomic charges (Fig. 2). 
$1s3s$ configuration weight function for $He$ (Fig. 3) in small $p$ region significantly 
exceeds $1s1s$ and $1s2s$ configuration weight functions, however, with the growth of $p$ $1s1s$ 
function dominating. For $Ar^{16+}$ $1s2s$ and $1s3s$ configuration weight functions are
similar to those of $He$, whereas their relative values in comparison with $1s1s$ function 
decrease significantly (Fig. 4). All configuration weight function exhibit
monotonic gradual changes with the increase in nuclear charge.

\paragraph*{Summery.}The proposed theory can be considered as an extension of configuration 
interaction method in which configuration weights 
depend on the values of the interaction potential, which makes the wave
function more flexible and eliminates the influence of the wave function
cusps on the convergence of the wave function to the exact one with an increase 
in a basis set. At the same time, the theory can be compared with explicitly
correlated methods since configuration weight functions explicitly depend on 
a particle-particle separation. The main difference between these theories is the 
form of the dependence which is prescribed in explicitly
correlated theories, whereas in the suggested theory it is  
obtained by the solution of the corresponding equations.

The solution of the model example with CI method and with the proposed
theory shows that the convergence of the proposed theory even in the lowest approximation is
faster than in CI method. In case of a full IPS $S_3$ or its approximation 
$S_{13}$ the exact energy values are rapidly reached with an increase in the basis set. 

The performed calculations of $He$-like ions show that the developed theory gives accuracy, at 
least, none the worse than the accuracy of the most precise techniques but with much less 
computational efforts. The use of only three configurations constructed from $1s$, $2s$, and $3s$ 
wave functions of non-interacting electrons in the nuclear field gives ground state energies of 
He-like ions lower than those of CI wave function with 35 configurations 
constructed from seven $s$, $p$, $d$, $f$, and $g$ Slater type orbitals \cite{Weiss}, and lower 
than those of configuration interaction wave function with 15 configuration constructed from 5 Slater 
orbitals and explicit $r_{12}$ terms up to 5 order \cite{Saha}, lower than Hylleraas-type wave function 
with more than three hundred terms. The results were obtained without iteration procedure of 
self-consistent field because the developed theory does not presuppose the use of the Hartree-Fock approximation as a preliminary step for precise calculations.

 The author gratefully acknowledges helpful discussions with the colleagues from Laboratory 
of Quantum Chemistry of Boreskov Institute of Catalysis. 

 \vspace{\baselineskip} 
\bibliographystyle{aipnum4-1}
%

\end{document}